\documentclass[twocolumn,showpacs,prb,superscriptaddress]{revtex4}
\newcommand{\figurewidth}{\columnwidth}
\usepackage{graphicx}
\newcommand{\chisg}{{\chi_{\rm SG}}}
\newcommand{\av}{_{\mathrm{av}}}

\begin{document}

\title{Spin glasses in the limit of an infinite number of spin components.}

\author{L.~W.~Lee}
\affiliation{Department of Physics,
University of California,
Santa Cruz, California 95064}

\author{A.~Dhar}
\affiliation{Raman Research Institute,
Bangalore 560080, India}

\author{A.~P.~Young}
\homepage[Homepage: ]{http://bartok.ucsc.edu/peter}
\email[Email: ]{peter@bartok.ucsc.edu}
\affiliation{Department of Physics,
University of California,
Santa Cruz, California 95064}


\begin{abstract}
We consider the spin glass model
in which the number of spin components, $m$, is
infinite. In the formulation of the problem appropriate for numerical
calculations proposed by several authors, we show that the order parameter
defined by the long-distance limit of the correlation functions is
actually zero and there is only ``quasi long range order'' below the
transition temperature. We also show that the spin glass transition
temperature is zero in three dimensions. 

\end{abstract}

\maketitle

\section{Introduction}
\label{sec:introduction}

It is of interest to study
a spin glass model in which the number of components of spin components $m$
is infinite,
because it provides some simplifications
compared with Ising ($m=1$) or Heisenberg ($m=3$) models. For example, in
mean field theory (i.e. for the infinite range model)
there is no ``replica symmetry breaking''~\cite{almeida:78b} so the
ordered state is characterized by a
single order parameter $q$, rather than by an infinite number
of order parameters (encapsulated in a
function $q(x)$) which are needed~\cite{parisi:80} for finite-$m$.

There has recently been renewed interest
~\cite{hastings:00,aspelmeier:04} in the $m=\infty$
model, and the interesting result
emerged from these studies that the \textit{effective} number of spin
components depends on the system size $N$ and is only really infinite in the
thermodynamic limit. One motivation for the present study is to investigate
some consequences of this result.

Further motivation for our present study comes from earlier work by two of
us~\cite{lee:03} which argued that
an isotropic vector spin glass ($m \ge
2$), as well as an Ising spin glass,
has a finite spin glass transition temperature $T_{SG}$ in three
dimensions. The results of Ref.~[\onlinecite{lee:03}] also indicate that
$T_{SG}$ is
very low compared with the mean field transition temperature,
$T_{SG}^{MF}$,
and \textit{decreases} with increasing $m$, see Table~\ref{TSG}. The results
in Table~\ref{TSG} suggest that $T_{SG}/T_{SG}^{MF}$ may be zero
in the $m=\infty$ limit in three dimensions,
and we investigate this possibility here.

\begin{table}
\begin{tabular*}{\figurewidth}{@{\extracolsep{\fill}} |l c |l l c| }
\hline
\hline
$m$  &  model  & $T_{SG}^{MF}$ & $T_{SG} $ & $T_{SG}/ T_{SG}^{MF}$  \\
\hline
  1 & Ising      & $2.45$ & 0.97~[\onlinecite{marinari:98,young:04}] & 0.40 \\
  2 & XY         & $1.22$ & 0.34~[\onlinecite{lee:03}] & 0.28 \\
  3 & Heisenberg & $0.82$ & 0.16~[\onlinecite{lee:03}] & 0.20 \\
\hline
\hline
\end{tabular*}
\caption{
Estimates of the spin glass transition temperature, relative to the mean field
value, $T_{SG}^{MF} = \sqrt{z}/m$, see Eq.~(\ref{TSGMF}),
for different values of $m$ for the
three-dimensional simple cubic lattice ($z=6$). The factor of $1/m$ in
$T_{SG}^{MF}$, appears because the spins
were normalized to unity in Refs.~[\onlinecite{lee:03,marinari:98,young:04}],
rather than to $m^{1/2}$ as
here. For the model used in this paper, $T_{SG}^{MF}$ is \textit{finite} for $M
\to \infty$.
\label{TSG}
}
\end{table}

In this paper, we study the $m=\infty$ SG model, both the infinite range
version and the short-range model
in three and two
dimensions. We find that we need to carefully specify the order
in which the limits $m \to \infty$ and the thermodynamic limit $N \to \infty$
are taken. In Ref.~\onlinecite{almeida:78b}, the $N \to \infty$ limit is taken
first (since a saddlepoint calculation is performed) and the $m \to \infty$
limit is taken at the end. However, in the formulation of the $m = \infty$
problem which has been proposed for numerical implementation in finite
dimensions~\cite{bray:82,morris:86,hastings:00,aspelmeier:04} and which we
use here, the limit $m\to\infty$ is taken first for a lattice of finite size.
In the latter case, we find that for $T < T_{SG}$ the spin glass correlations
decay with a power of the distance $r$ and tend to zero for $r \to \infty$, so
the order parameter, defined in terms of the long-distance limit of the
correlation function, is actually zero. Nonetheless, there can still be a
transition at $T_{SG}$ separating a high temperature phase where the
correlations decay exponentially, from the the low temperature phase where
they decay with a power law.
By contrast, if one takes $N \to \infty$ first with
$m$ finite, the power law decay eventually changes to a constant at large $r$
and so a non-zero spin glass order parameter can be defined, as in
Ref.~\onlinecite{almeida:78b}.

We give phenomenological arguments for these
conclusions and back them up (for the case where $m\to\infty$ is taken first)
by numerical results at zero temperature.  We also find, from numerical
results at finite temperature, that $T_{SG}/T_{SG}^{MF} = 0$ in three
dimensions for $m=\infty$, consistent with the trend of the results in
Table~\ref{TSG}.

In Sec.~\ref{sec:model} we discuss the model and the methods used to study it
numerically. In Sec.~\ref{sec:T0} we describe our results at $T=0$ for both
short-range and the infinite-range model, while in
Sec.~\ref{sec:finiteT} we describe finite temperature results for short-range
models. Our
conclusions are summarized in Sec.~\ref{sec:conclusions}. 

\section{Model and method}
\label{sec:model}
We take the Edwards-Anderson~\cite{edwards:75} Hamiltonian
\begin{equation}
{\cal H} = \sum_{\langle i, j \rangle} J_{ij} \mathbf{S}_i \cdot \mathbf{S}_j
\, , 
\label{ham}
\end{equation}
where the spins $\mathbf{S}_i$ ($ i = 1, \cdots, N$) are classical vectors
with $m$ components and normalized to length $m^{1/2}$, i.e. $\mathbf{S}_i^2 =
m$. As we shall see,
this normalization is necessary to get a finite transition temperature in
the mean field limit. The $J_{ij}$ are independent random variables with a
Gaussian distribution with zero
mean. We consider both the infinite range model and short-range models with
nearest-neighbor interactions in two and three dimensions.
For the infinite range model, the standard deviation is taken to be
$1/\sqrt{N-1}$ 
while for the short-range
models the standard deviation is set to be unity. According
to the mean field
approximation, the spin glass transition temperature is
\begin{equation}
T_{SG}^{MF} = { \langle \mathbf{S}_i^2 \rangle \over m} \left[\sum_j
J_{ij}^2 \right]\av^{1/2}
\, ,
\label{TSGMF}
\end{equation}
where $[\cdots]\av$ indicates an average over the disorder. Hence,
for the infinite range model, (where mean field theory is exact)
Eq.~(\ref{TSGMF})
gives $T_{SG} = T_{SG}^{MF} = 1$, while for the short-range case it gives
$T_{SG}^{MF} = \sqrt{z}$, where $z$ is the number of nearest neighbors (4 for
the square lattice and 6 for the simple cubic lattice).

As shown in other work~\cite{bray:82,morris:86,hastings:00,aspelmeier:04},
the problem can be simplified
for $m=\infty$. The spin-spin correlation function,
\begin{equation}
C_{ij} \equiv {1 \over m} \langle \mathbf{S}_i \cdot \mathbf{S}_j \rangle ,
\label{Cij}
\end{equation}
is given by
\begin{eqnarray}
T^{-1} C_{ij} & = & \left(A^{-1}\right)_{ij},  \qquad \mbox{where}
\label{chi} \\
A_{ij} & = & H_i \delta_{ij} - J_{ij} ,
\label{A}
\end{eqnarray}
and the $H_i$ have to be determined self consistently to enforce (on average)
the length constraint on the spins,
\begin{equation}
C_{ii} = 1 \, .
\label{self_cons}
\end{equation}
Angular brackets, $\langle \cdots \rangle$, refer to a thermal average for a
given set of disorder. Eq.~(\ref{self_cons}) with $i=1,\cdots,N$
represents $N$ equations 
which have to be solved for the $N$ unknowns $H_i$. In
Sec.~\ref{sec:finiteT} we will solve these equations numerically for a range of
sizes at finite temperature. We emphasize that in
Eqs.~(\ref{Cij})--(\ref{self_cons}) the limit $m \to \infty$ has
been taken with $N$ finite. This is the opposite order of limits from that in
the analytical work of Ref.~\onlinecite{almeida:78b} where $N \to \infty$ was
taken before $m \to \infty$. As we shall see, the results from
the two orders of
limits are different.

\begin{center}
\begin{figure}
\includegraphics[width=\figurewidth]{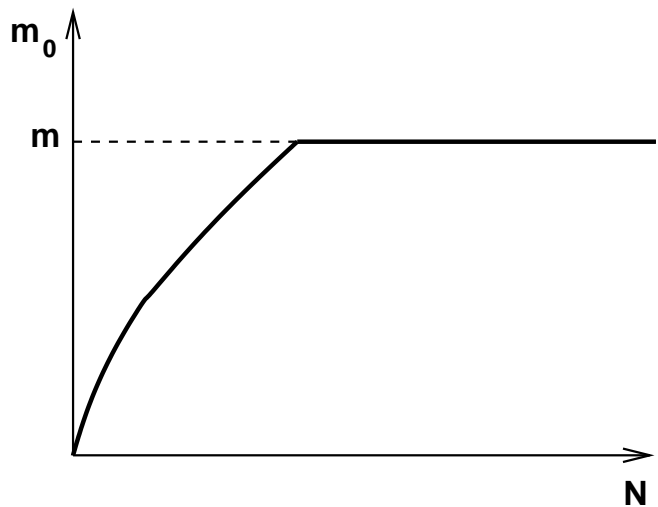}
\caption{
A plot of the effective number of spin components as a function of system size
$N$. For small $N$, $m_0 \sim N^\mu$, but once $m_0$ hits the actual number of
spin components $m$, it sticks at $m$ as $N$ is further increased.
}
\label{fig:m0}
\end{figure}
\end{center}

Eqs.~(\ref{Cij})--(\ref{self_cons}) are not well defined at $T=0$. However,
Aspelmeier and Moore~\cite{aspelmeier:04}
pointed out that one can solve the $m=\infty$ problem
\textit{directly} at $T=0$, using the following method.  At zero temperature
there are no thermal fluctuations so each spin lies parallel to its local
field, i.e.
\begin{equation}
\textbf{S}_i = H_i^{-1} \sum_j J_{ij} \textbf{S}_j \, ,
\label{self_cons_T0}
\end{equation}
where $m^{1/2} H_i$ is the magnitude of the local field on site $i$.
Remarkably, it was shown by Hastings~\cite{hastings:00} that these local
fields are precisely the zero temperature limit of the $H_i$ in Eq.~(\ref{A}).
Hastings~\cite{hastings:00} also showed that the number of independent spin
components which are non-zero in the ground state, which we call $m_0$, cannot
be arbitrarily large, but satisfies the bound
\begin{equation}
m_0 < \sqrt{2 N} \, . 
\label{m0_bound}
\end{equation}
This means that one can always perform a global
rotation of the spins such that only $m_0$ components have a non-zero
expectation value and the remaining $m - m_0$ components vanish. Thus one can
think of $m_0$ as the \textit{effective number of spin components}. 
If $m$ is finite, then, at some value of $N$,
$m_0$ would equal the actual number of spin components
$m$. At this point,
all spin components are used so
$m_0$ ``sticks'' at the value $m$ as $N$ is further increased, see
Fig.~\ref{fig:m0}.

More generally we can write Eq.~(\ref{m0_bound}) as
\begin{equation}
m_0 \sim N^\mu , \quad (m_0 < m)
\label{mu}
\end{equation}
and the bound in Eq.~(\ref{m0_bound})
gives $\mu \le 1/2$. Later, we will determine $\mu$ numerically
for several models.
For Eqs.~(\ref{Cij})--(\ref{self_cons}) to be valid we need
$m > m_0$ which corresponds to the curved part of the line in
Fig.~\ref{fig:m0}. As discussed above, this corresponds to taking
the limit $m \to \infty$ \textit{first}, followed
by the limit $N \to \infty$.
Since $m_0$ increases with $N$ one needs larger values
of $m$ for larger lattice sizes. This will be important in what follows. 

We therefore see that we can numerically
solve the $m=\infty$ problem at $T=0$ on a finite lattice by taking a
number of spin components which is \textit{finite} but greater than $m_0$, and
solving Eqs.~(\ref{self_cons_T0}). To do this we cycle through the lattice,
and
at site $i$, say, we calculate $H_i$ from
\begin{equation}
H_i  ={1\over m^{1/2}} \left| \sum_j J_{ij} \mathbf{S}_j\right| \, .
\label{Hi}
\end{equation}
We then set $\mathbf{S}_i$
to the value given by Eq.~(\ref{self_cons_T0}) so it lies parallel to its
instantaneous local field. This is
repeated for each site $i$, and then the whole procedure iterated 
iterated to convergence.
Although spin glasses with finite-$m$ have many
solutions of Eqs.~(\ref{self_cons_T0}), it turns out that
for $m=\infty$ (in practice this means $m >
m_0$) there is a unique stable solution~\cite{bray:81}, so the numerical
solution of Eqs.~(\ref{self_cons_T0}) is straightforward. We will discuss
our numerical results at $T=0$ using Eqs.~(\ref{self_cons_T0})
in Sec.~\ref{sec:T0}, and here we simply note that we do indeed find a unique
solution of these equations.

Next we consider the order parameter in spin glasses for $m=\infty$. In the
absence of a symmetry breaking field, one defines the long range order
parameter, $q$, by the behavior of the spin-spin correlation function
$[C_{ij}^2]\av$ at large distances, i.e.
\begin{equation}
q^2 = \lim_{R_{ij} \to \infty} [ C_{ij}^2 ]\av \, \qquad
\mbox{(short\ range),}
\label{q2sr}
\end{equation}
where $R_{ij} = |\mathbf{R}_i - \mathbf{R}_j|$.
For the infinite-range model, any distinct pair of sites will do, and so
\begin{equation}
q^2 = [ C_{ij}^2 ]\av \ (i \ne j) \qquad
\mbox{(infinite\ range).}
\label{q2ir}
\end{equation}

We now give phenomenological arguments, which will be supported by numerical
data in Sec.~\ref{sec:T0}, that $q$ obtained from Eqs.~(\ref{q2sr})
and (\ref{q2ir}),
in which $C_{ij}$ is determined by
Eqs.~(\ref{Cij})--(\ref{self_cons}), is actually \textit{zero} for $m=\infty$,
and that, at best, spin correlations have only ``quasi-long range order''. For
the short range case, this means that $[C_{ij}^2]\av$ decays with a power
of the distance $R_{ij}$, while for the infinite range
model the correlation function in Eq.~(\ref{q2ir}) tends to zero with a power
of $N$.

To see why this is the case, we take $T=0$ and
consider first the infinite-range model. For a
given $N$, the spins ``splay out'' in $m_0 \sim N^\mu$ directions. We
expect the spins to point, on average, roughly equally in all directions in this
$m_0$-dimensional space. Now $C_{ij}$ in Eq.~(\ref{Cij}) is equal to $\cos
\theta_{ij}$ where $\theta_{ij}$ is the angle between $\mathbf{S}_i$ and
$\mathbf{S}_j$. We take the square and average equally over all
directions. To do the average, take a coordinate system with the polar axis
along $\mathbf{S}_i$, so $\theta_{ij} = \theta_j$ the polar angle of
$\mathbf{S}_j$. Then we have
\begin{eqnarray}
q^2 & = & [ C_{ij}^2 ]\av = \langle \cos^2 \theta_j \rangle \nonumber \\
& = & {1 \over \mathbf{S}^2} \langle S_z^2\rangle 
\sim {1 \over m_0} {\sum_{\alpha = 1}^{m_0} \langle S_\alpha^2\rangle \over
\mathbf{S}^2}
= {1 \over m_0} \sim N^{-\mu} \, ,
\label{q2}
\end{eqnarray}
where we used the result that the average is roughly
the same for all the $m_0$ spin
components.  Since $\mu$ will turn out to be non-zero
it follows that \textit{the order parameter tends
to zero} with a power of the size of the system. The same will be true at
temperatures $T < T_{SG}$, while above $T_{SG}$ the order parameter as defined
here will vanish faster, as $1/N$.

How can we reconcile this vanishing order parameter
with earlier results~\cite{almeida:78b} that
the order parameter is non zero below $T_{SG} = 1$, and 
in particular is unity at $T=0$. The difference
comes in part because $q^2$ in Ref.~[\onlinecite{almeida:78b}], which we call
$q^2_{\rm AJKT}$, is $m$ times
our $q^2$,
and so
\begin{equation}
q^2_{\rm AJKT} = m q^2 \sim  {m \over m_0}, \quad (T = 0) .
\label{q2AJKT}
\end{equation}
The other difference is that
Ref.~[\onlinecite{almeida:78b}] performs the limit $N \to \infty$ first, 
which corresponds to being on the horizontal part of the line in
Fig.~\ref{fig:m0}, so
$m_0 = m$. From Eq.~(\ref{q2AJKT}) we then
get $q^2_{\rm AJKT} = \text{const.}$ at $T=0$ in agreement
with Ref.~[\onlinecite{almeida:78b}].

Going back to
the calculation of $C_{ij}$, if one
sums $[C_{ij}^2]\av$ for the infinite range model
over all pairs of sites we find that the spin glass
susceptibility $\chisg$ at $T=0$ is given by
\begin{equation}
\chisg ={1 \over N} 
\sum_{i,j} [ C_{ij}^2]\av = 1 + (N-1)q^2 \simeq N q^2 \sim N^{1 -
\mu} \, .
\label{chisg}
\end{equation}

Turning now to the short-range case, we expect that $\chisg  \sim N^{1 - \mu} $
will still be true, which implies that correlations decay with a power of
distance. Assuming that $[ C_{ij}^2]\av \sim 1/R_{ij}^y$
for some exponent $y$, then integrating
over all $\mathbf{r}$ up to $r = L$ (where $N=L^d$) and requiring that the
result goes as $N^{1 - \mu} $, gives $y = d\, \mu$, i.e.
\begin{equation}
[ C_{ij}^2]\av \sim {1 \over R_{ij}^{d\, \mu} } \, .
\label{Cij_sr}
\end{equation}
Such power law decay
is often called ``quasi long range
order''. We expect that Eq.~(\ref{Cij_sr}) will be true quite generally at
$T=0$ and everywhere below $T_{SG}$ if $T_{SG} > 0$. Note that this implies
that $q=0$ according to Eq.~(\ref{q2sr}). Above $T_{SG}$,
$[C_{ij}^2]\av$ will decay to zero exponentially with distance.

If $m$ is large but finite, then $[C_{ij}^2]\av$ will saturate when
$R_{ij}$ is sufficiently large
that all the spin components are used. This happens when 
$[C_{ij}^2]\av \sim 1/m$, i.e. for $R_{ij} \gtrsim m^{1/d \, \mu}$. In this
case, $q^2_{\rm AJKT} = m q^2$ will be finite according to Eq.~(\ref{q2sr}).

\begin{table}
\begin{tabular}{| r | r |}
\hline
\hline
$N$  &  $N_{\rm samp}$  \\
\hline
  32 & $1000$ \\
  64 & $1000$ \\
 128 & $1000$ \\
 256 & $1000$ \\
 512 & $1000$ \\
1024 & $ 777$ \\
2048 & $ 302$ \\
\hline
\hline
\end{tabular}
\caption{
Number of samples used in the $T=0$ studies of the infinite-range model.
\label{table:SK}
}
\end{table}

\begin{center}
\begin{figure}
\includegraphics[width=\figurewidth]{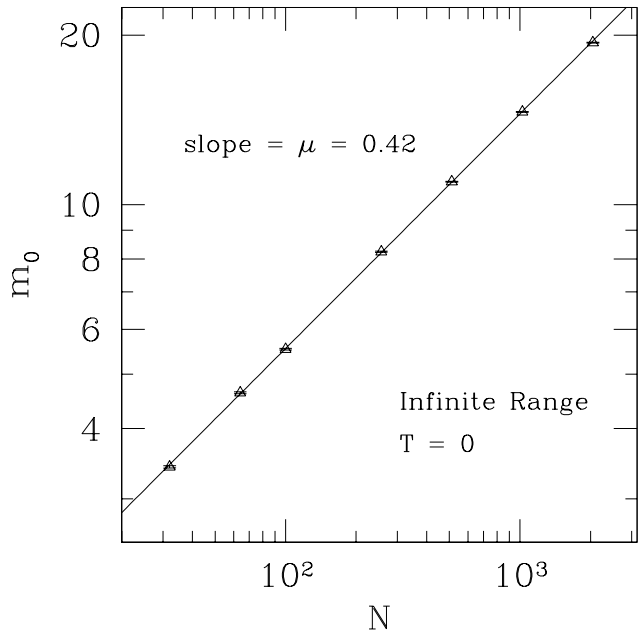}
\caption{
The average number of non-zero spin components in the ground state, $m_0$,
as a function of $N$ for the infinite-range model. We see that $m_0$ increases
like $N^\mu$ with $\mu$ close to $2/5$ as expected.
}
\label{fig:m0_ir}
\end{figure}
\end{center}
\begin{center}
\begin{figure}
\includegraphics[width=\figurewidth]{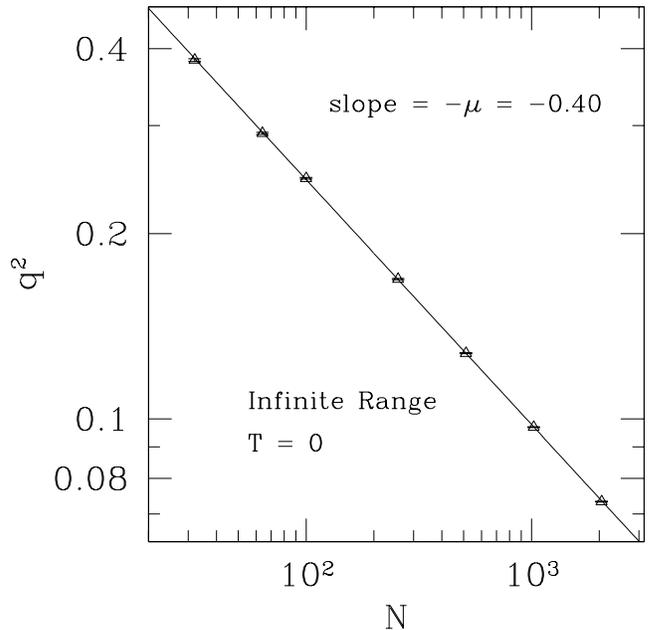}
\caption{
The square of the order parameter at $T=0$ for infinite range model. As
expected it decreases like $N^{-\mu}$ with $\mu = 2/5$.
}
\label{fig:q2_ir}
\end{figure}
\end{center}

In Secs.~\ref{sec:T0_IR} and \ref{sec:T0_sr} we will provide numerical support
for Eq.~(\ref{chisg}) for the infinite-range and short-range cases
respectively.

\section{Results at zero temperature}
\label{sec:T0}

\subsection{Infinite Range Model}
\label{sec:T0_IR}

We consider a range of 
lattice sizes up to $N=2048$ and for each size the number of
samples is shown in Table~\ref{table:SK}.

The average
number of non-zero spin components in the ground state is given by
Eq.~(\ref{mu}), for which it has been shown
that\cite{aspelmeier:04,hastings:00}
\begin{equation}
\mu = 2/5 \, , \qquad \mbox{(infinite\ range.)}
\end{equation}
exactly. This result has been confirmed numerically~\cite{aspelmeier:04}. 
Our results for $\mu$ are shown in Fig.~\ref{fig:m0_ir} and indeed give $\mu$
close to 2/5. The small deviation is presumably due to corrections to scaling.

We also calculated $q^2$ at $T=0$ from Eq.~(\ref{q2ir}). In Eq.~(\ref{Cij}) the
thermal average, $\langle \cdots \rangle$, is unnecessary, and the spin
directions are determined by solving Eqs.~ (\ref{self_cons_T0}) and
(\ref{Hi}).
The results for are shown in Fig.~\ref{fig:q2_ir},
showing that it vanishes with exponent $-\mu$ as a
function of $N$, as expected from Eq.~(\ref{q2}).

\begin{center}
\begin{figure}[tb!]
\includegraphics[width=\figurewidth]{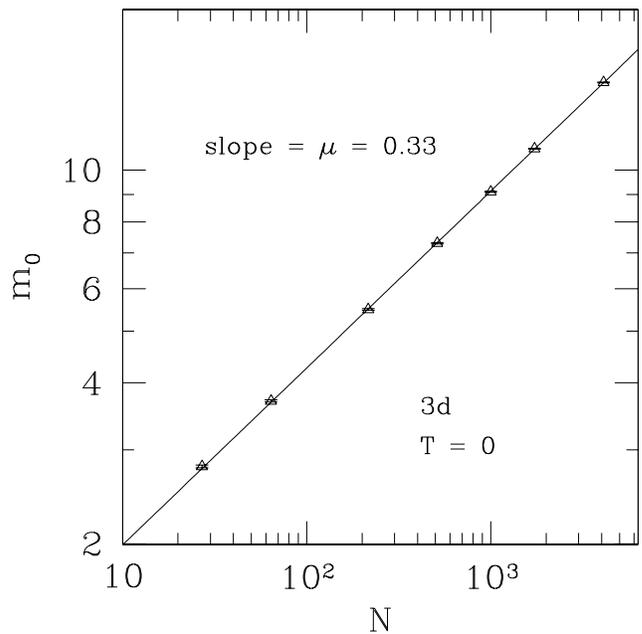}
\caption{
The average number of non-zero spin components in the ground state, $m_0$,
as a function of $N$ for the short-range model in $d=3$.
We see that $m_0$ increases
like $N^\mu$ with $\mu \simeq 0.33$.
}
\label{fig:m0_3d}
\end{figure}
\end{center}

\begin{table}
\begin{tabular}{| r | c | c | c |}
\hline
\hline
 & \multicolumn{2}{| c |}{$T = 0$} & $T > 0 $ \\
\hline
$L$  &  $N_{\rm samp} \ (m_0)$ & $N_{\rm samp} \ (\chisg)$ & $N_{\rm samp}$ \\
\hline
  3 & $1000$ & --     & --      \\
  4 & $1000$ & $1000$ & $100$   \\
  6 & $1000$ & $1000$ & $100$   \\
  8 & $1000$ & $1000$ & $100$   \\
 10 & $1000$ & --     & --      \\
 12 & $1105$ & $1105$ & $100$   \\
 16 & $ 785$ & $ 785$ & --     \\
 24 & --     & $ 500$ & --     \\
\hline
\hline
\end{tabular}
\caption{
Number of samples used in the calculations 
for the short-range model in three-dimensions. 
\label{table:3d}
}
\end{table}

\begin{center}
\begin{figure}[tb!]
\includegraphics[width=\figurewidth]{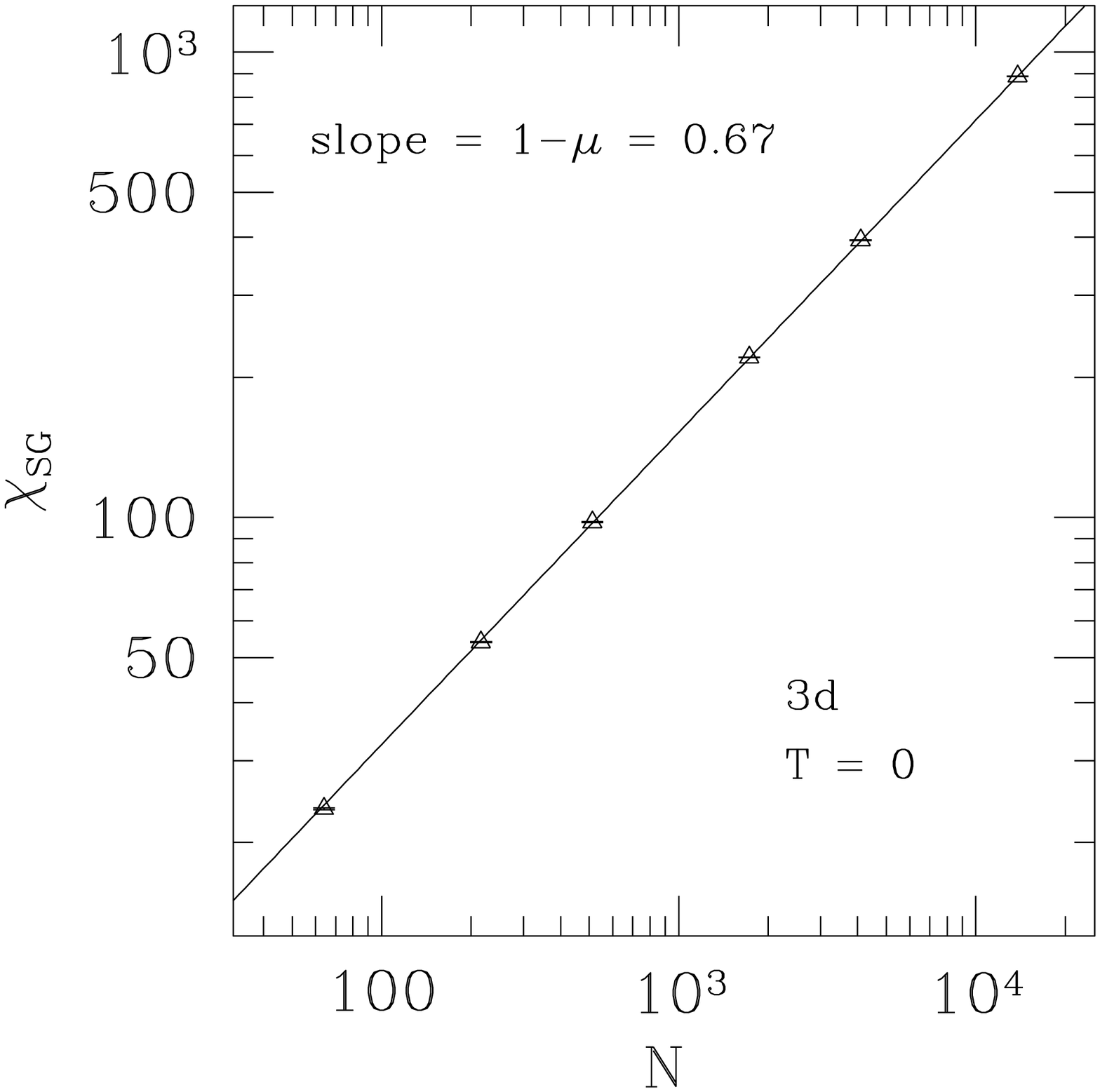}
\caption{
The spin glass susceptibility for the short-range model in $d=3$ for different
system sizes. As expected it varies as $N^{1-\mu}$, where $\mu \simeq 0.33$
was also found in Fig.~\ref{fig:m0_3d}.
}
\label{fig:chisg0_3d}
\end{figure}
\end{center}

\begin{table}
\begin{tabular}{| r | c | c | c |}
\hline
\hline
& \multicolumn{2}{| c |}{$T=0$} & $T > 0$ \\
\hline
$L$  &  $N_{\rm samp} \ (m_0)$ & $N_{\rm samp} \ (\chisg)$ & $N_{\rm samp}$  \\
\hline
  4 & $1000$ & $1000$ & $1000$   \\
  6 & $1000$ & $1000$ & $1000$   \\
  8 & $1000$ & $1000$ & $1000$   \\
 10 & $1000$ & --     & --       \\
 12 & $1000$ & $1000$ & $1000$   \\
 14 & $1000$ & --     & --       \\
 16 & $1000$ & $1000$ & $1000$  \\
 18 & $1000$ & --     & --      \\
 20 & $1000$ & --     & --      \\
 22 & $1000$ & --     & --      \\
 24 & --     & $1000$ & $ 500$  \\
 28 & $1000$ & --     & --      \\
 32 & $1000$ & $1000$ & $ 309$  \\
 48 & --     & $ 472$ & $ 136$  \\
 64 & $1016$ & $1016$ & --      \\
\hline
\hline
\end{tabular}
\caption{
Number of samples used in the calculations 
for the short-range model in two-dimensions. 
\label{table:2d}
}
\end{table}
\begin{center}
\begin{figure}[tb!]
\includegraphics[width=\figurewidth]{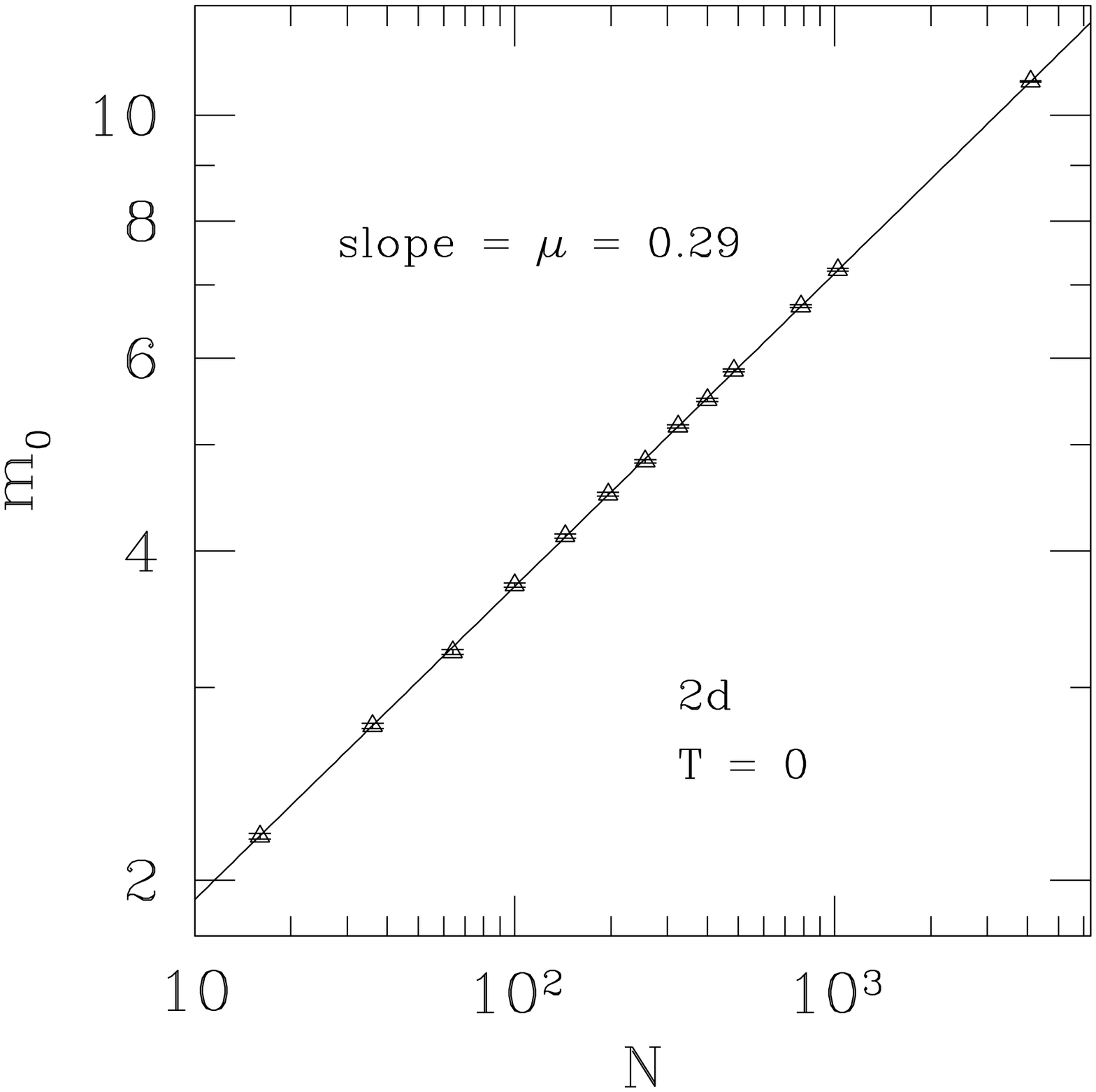}
\caption{
The average number of non-zero spin components in the ground state, $m_0$,
as a function of $N$ for the short-range model in $d=2$.
We see that $m_0$ increases
like $N^\mu$ with $\mu \simeq 0.29$.
}
\label{fig:m0_2d}
\end{figure}
\end{center}

\begin{center}
\begin{figure}[tb!]
\includegraphics[width=\figurewidth]{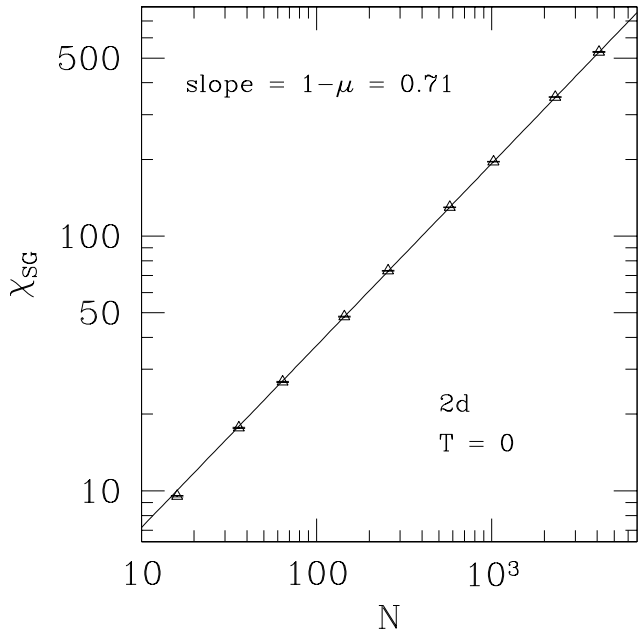}
\caption{
The spin glass susceptibility for the short-range model in $d=2$ for different
system sizes. As expected it varies as $N^{1-\mu}$, where $\mu \simeq 0.29$
was also found in Fig.~\ref{fig:m0_2d}.
}
\label{fig:chisg0_2d}
\end{figure}
\end{center}

\subsection{Short Range models}
\label{sec:T0_sr}
First of all we describe our results for three dimensions. The number of
samples is shown in Table~\ref{table:3d}.

Our results for $\mu$ are shown in Fig.~\ref{fig:m0_3d}, indicating that $\mu
\simeq 0.33$, definitely different from the infinite range result of $2/5$.
The results for $\chisg$ as a function of $N$
are shown in Fig.~\ref{fig:chisg0_3d}.
We see that $\chisg$ grows with an exponent $1-\mu$ with the same value of
$\mu$ as in Fig.~\ref{fig:m0_3d}.
We therefore find that $d\, \mu \simeq 1.0$, and so,
from Eq.~(\ref{Cij_sr}), the spin glass correlations decay as
\begin{equation}
[C_{ij}]^2\av \sim {1 \over R_{ij}} , \qquad (d = 3, \ T = 0) \, .
\end{equation}
(It is of course possible that power of $R_{ij}$ may not be exactly $-1$.)

Next we describe our results for two dimensions. The number of samples used is
shown in Table~\ref{table:2d}.
Our results for $\mu$ are shown in Fig.~\ref{fig:m0_2d}, and give $\mu \simeq
0.29$.
The data for $\chisg$ is shown in Fig.~\ref{fig:chisg0_2d}. We see that
$\chisg$ increases as $N^{1-\mu}$ with the same $\mu$ as determined from
Fig.~\ref{fig:m0_2d}. 
We therefore find that $d\, \mu \simeq 0.58$, and so,
from Eq.~(\ref{Cij_sr}), the spin glass correlations decay as
\begin{equation}
[C_{ij}]^2\av \sim {1 \over R_{ij}^{0.58}} , \qquad (d = 2, \ T = 0) \, .
\end{equation}

\section{Results for short range models at finite temperature}
\label{sec:finiteT}

\begin{center}
\begin{figure}[tb!]
\includegraphics[width=\figurewidth]{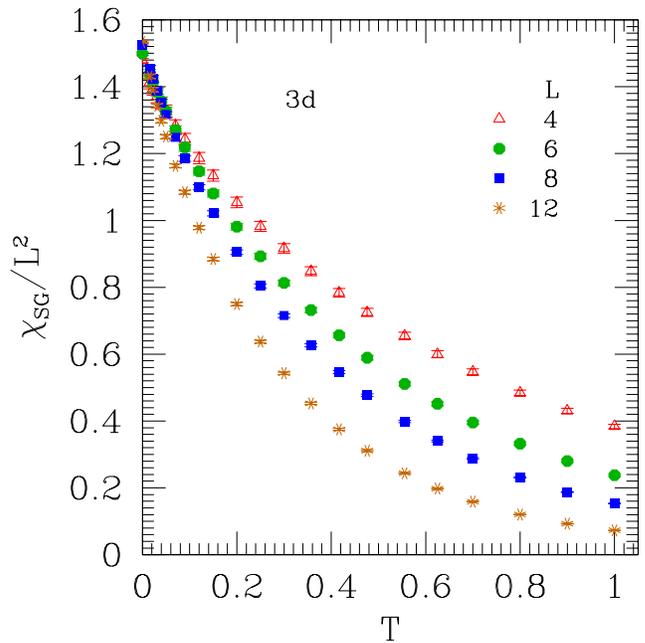}
\caption{
The spin glass susceptibility as a function of temperature in three
dimensions. The vertical axis has been divided by $L^{d(1-\mu)}$, in which we
took $\mu = 1/3$ in order to collapse the data at $T=0$
according to the data in Figs.~\ref{fig:m0_3d} and
\ref{fig:chisg0_3d}.
}
\label{fig:chisgT_3d}
\end{figure}
\end{center}

\begin{center}
\begin{figure}[tb!]
\includegraphics[width=\figurewidth]{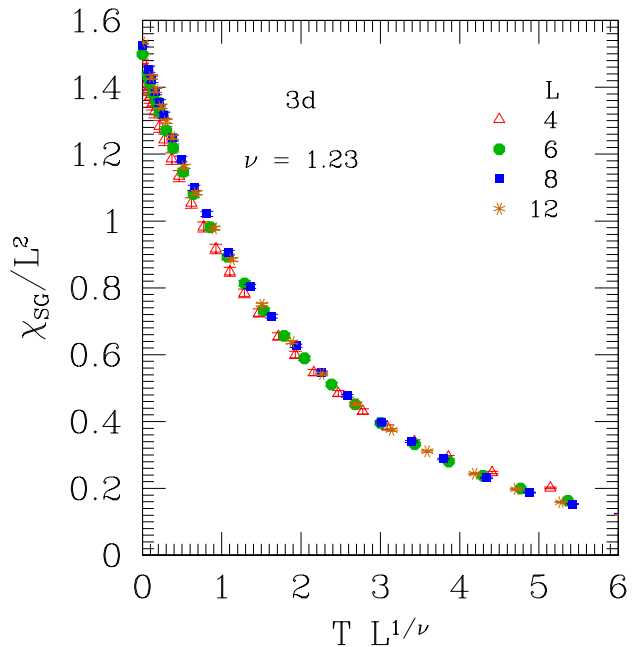}
\caption{
A scaling plot of the spin glass susceptibility in Fig.~\ref{fig:chisgT_3d}
assuming a zero temperature transition.
}
\label{fig:chisgT_3d_scale}
\end{figure}
\end{center}
We have determined finite temperature properties by solving
Eqs.~(\ref{chi})--(\ref{self_cons})
self-consistently using the Newton-Raphson method. We start
at high temperature, $T = T_1$ say, and take
our initial guess to be $H_i = 1/\beta$ which is the solution obtained
perturbatively to first order in $1/T$.
We then solve the equations at
successively lower temperatures, $T_1 > T_2 > T_3 > T_4 \cdots$, and obtain the
initial guess for the $H_i$ at temperature $T_{i+1}$ by integrating
the equations~\cite{aspelmeier:04}
\begin{equation}
{d H_i \over d \beta} = - \sum_j \left(B^{-1}\right)_{ij} \, ,
\end{equation}
in which
\begin{equation}
B_{ij} = (\beta C_{ij})^2 \, ,
\end{equation}
from $\beta_i$ to $\beta_{i+1}$ ($\beta = 1/T$).

Results for $\chisg$
in $d=3$ are shown in
Fig.~\ref{fig:chisgT_3d},
in which we scaled the vertical
axis by $L^{d(1-\mu})\ (= L^2)$ so the data collapses at $T = 0$.
If we assume a zero temperature transition, the data should fit the
finite-size scaling form
\begin{equation}
\chisg = L^{d(1-\mu)} X \left(L^{1/\nu} T\right) \, .
\label{chiscale}
\end{equation}
where $X(x) \to \text{const.}$ for $x \to 0$, and
the power law prefactor in front of the scaling function $X(x)$ then
gets the $T=0$ limit correct.
Figure ~\ref{fig:chisgT_3d_scale} shows an appropriate
scaling plot with $\nu = 1.23$. Apart from the smallest size, $L=4$,
the data clearly collapses well. By
considering different values of $\nu$ we estimate
\begin{equation}
\nu = 1.23 \pm 0.13 \qquad (d=3) \, .
\end{equation}
This result can be compared with that of Morris et
al.~\cite{morris:86} who quote $\nu = 1.01 \pm 0.02$. Since our
results cover a larger range of sizes and have better statistics, we feel that
the error bars of Morris et al.~are too optimistic. Assuming this,
our result is consistent with theirs. 

\begin{center}
\begin{figure}[tb!]
\includegraphics[width=\figurewidth]{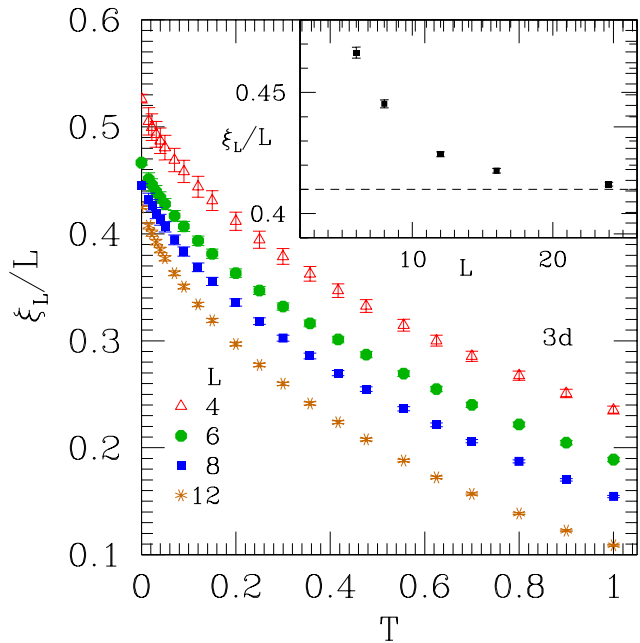}
\caption{
The main figure is a plot of $\xi_L/L$ against $T$ in three dimensions.  The
inset shows $\xi_L/L$ at $T=0$ as a function of $L$. The dashed line is a
guide to the eye.
}
\label{fig:xiT_and_0_3d}
\end{figure}
\end{center}

We should, however, also test to see if the data can be fitted with a
finite value for $T_{SG}$. To do
this, it is convenient to analyze the correlation
length of the finite system, $\xi_L$, and plot the dimensionless ratio
$\xi_L/L$ which has the expected scaling form~\cite{ballesterosetal:00,lee:03}
\begin{equation}
{\xi_L \over L} = F\left( L^{1/\nu} (T - T_{SG})\right) \, 
\end{equation}
without any unknown power of $L$ multiplying the scaling function $F$. Hence
the data for different sizes should intersect at $T_{SG}$ and also splay out
below $T_{SG}$. To determine $\xi_L$ we Fourier transform $[C_{ij}^2]\av$ to
get $\chisg({\bf k})$ and then use~\cite{ballesterosetal:00,lee:03}
\begin{equation}
\xi_L = {1 \over 2 \sin(k_{\rm min}/2)} \left( {\chisg(0) \over
\chisg({\bf k}_{\rm min})} - 1 \right)^{1/2} \, ,
\end{equation}
where ${\bf k}_{\rm min} = (2\pi/L)(1, 0, 0)$ is the smallest non-zero
wavevector on the lattice.

The results are shown in the main part of
Fig.~\ref{fig:xiT_and_0_3d}. The data don't intersect 
at any temperature, but seem to be approaching an intersection at $T=0$ for
the larger sizes. To test out this possibility,
we have computed the correlation length
directly at
$T=0$, from the solution of Eqs.~(\ref{self_cons_T0}) and (\ref{Hi}),
where we can study larger sizes than in the finite-$T$ formulation of
Eqs.~(\ref{Cij})--(\ref{self_cons}).  The data is shown in the inset to
Fig.~\ref{fig:xiT_and_0_3d}. It indicates, fairly convincingly, that $\xi_L/L$
approaches a constant for $L\to\infty$ at $T=0$,
and hence that there is a transition at $T=0$.

\begin{center}
\begin{figure}[tb!]
\includegraphics[width=\figurewidth]{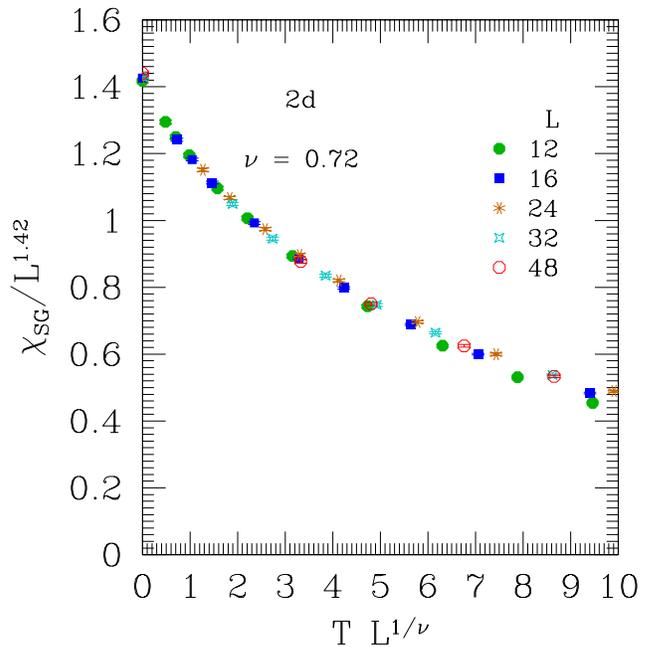}
\caption{
A scaling plot of the data for $\chisg$ in two dimensions, assuming a zero
temperature transition. In the vertical axis, $\chisg$ is divided by
$L^{d(1-\mu)} \simeq L^{1.42}$ so that the data collapses at $T=0$.
}
\label{fig:chisgT_2d_scale}
\end{figure}
\end{center}

In $d=2$ it is well established that $T_{SG}= 0$ even for the Ising case. 
A scaling plot for $\chisg$ for $m=\infty$
in $d=2$, corresponding to Eq.~(\ref{chiscale}), is shown in
Fig.~\ref{fig:chisgT_2d_scale} with $\nu = 0.72$, which gives the best data
collapse for larger sizes, and $d(1- \mu) = 1.42$ which
is obtained from the $T=0$ results in Sec.~\ref{sec:T0}. Again the data scales
well.  Overall we estimate
\begin{equation}
\nu = 0.72 \pm 0.05 \qquad (d=2, \quad \text{from } \chisg) \, .
\label{nu_2d_chisg}
\end{equation}
This is consistent with
the results in Morris et al.~\cite{morris:86} who quote $\nu = 0.65 \pm 0.02$.

\begin{center}
\begin{figure}[tb!]
\includegraphics[width=\figurewidth]{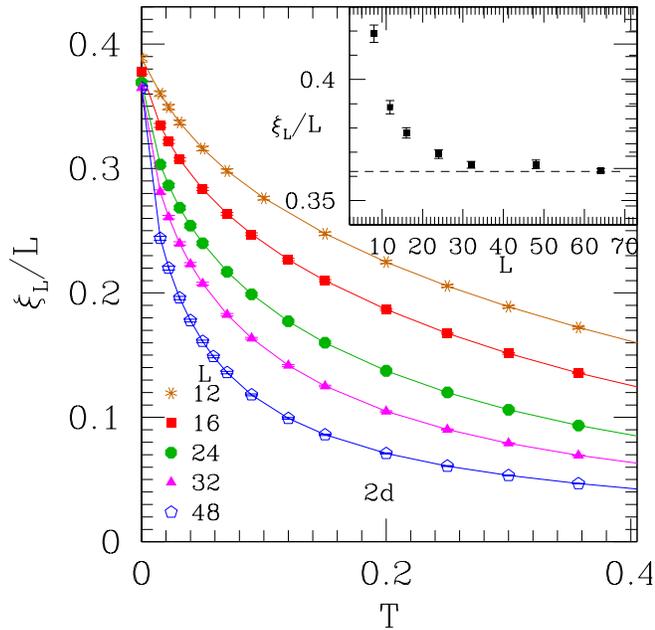}
\caption{
Data for $\xi_L/L$ as a function of $T$
in two dimensions. Clearly the data for larger sizes is
merging at $T=0$ indicating a transition at $T_{SG}=0$.
The inset shows data for $\xi_L/L$ at $T=0$ confirming that the data becomes
independent of size at $T=0$. The dashed line is a guide to the eye.
}
\label{fig:xiT_and_0_2d}
\end{figure}
\end{center}

\begin{center}
\begin{figure}[tb!]
\includegraphics[width=\figurewidth]{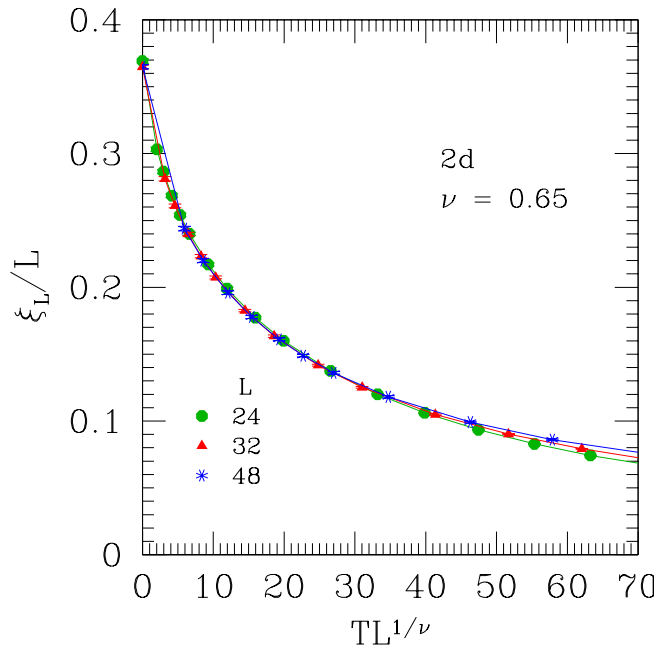}
\caption{
A scaling plot of the data from the largest sizes 
for $\xi_L/L$ in two dimensions assuming $T_{SG} = 0$.
}
\label{fig:xiT_scale_2d}
\end{figure}
\end{center}

We have also computed the correlation length $\xi_L/L$ in two dimensions, and
show the data in Fig.~(\ref{fig:xiT_and_0_2d}). The curves become independent
of size, for large $L$, at $T=0$, confirming that $T_{SG} = 0$. A scaling plot
of the data for the largest sizes ($L \ge 24$) in Fig.~\ref{fig:xiT_scale_2d}
has the best data collapse with $\nu = 0.65$ and altogether we estimate
\begin{equation}
\nu = 0.65 \pm 0.05 \qquad (d=2, \quad \text{from } \xi_L/L) \, ,
\end{equation}
which is consistent with our estimate from $\chisg$
in Eq.~(\ref{nu_2d_chisg}), and
with the result of Morris et al.~\cite{morris:86}.

\section{Conclusions}
\label{sec:conclusions}

We have considered the spin glass in the limit where the spins have an
infinite number of components. In the formulation of this problem appropriate
for numerical calculations~\cite{bray:82,morris:86,hastings:00,aspelmeier:04},
where the limit $m\to\infty$ is taken with $N$ finite, we find that
the order parameter, defined in terms of correlation functions
in zero (symmetry-breaking) field, vanishes. Instead, below $T_{SG}$,
there is only ``quasi-long range order'' in which the correlations
decay to zero with a power of distance.  Whereas we define the order parameter
in terms of the the \textit{long distance limit} of the correlation functions,
Aspelmeier and Moore~\cite{aspelmeier:04} define a \textit{local} order
parameter in terms of the contribution to the constraint in
Eq.~(\ref{self_cons}) that comes from the eigenmodes with zero eigenvalue of
the matrix $A_{ij}$. They argue their order 
parameter is related to the susceptibility in the
presence of a small field $h$, where the limit $N\to\infty$ is taken before
the limit $h \to 0$ in order to break the symmetry. From numerics on the
infinite-range model, Aspelmeier
and Moore claim that their order parameter agrees with that of Almeida et
al.~\cite{almeida:78b}. 

However, in a sensible physical model, \textit{any}
reasonable definition of the order
parameter should give the same answer. In particular, one should be able to
obtain the square of the order parameter
from the long distance limit of the correlation
function (off-diagonal long range order) in zero field, and get the same
answer as the local expectation value of the spin in the presence of a small
symmetry breaking field. This does not appear to be the case for the
$m=\infty$ model if the limit $m \to \infty$ is taken before $N \to \infty$.

On the other hand, if the thermodynamic limit, $N \to \infty$, is taken with
$m$ large but finite, then the correlations saturate at a value of order $1/m$
at large distance, and so a finite spin glass order parameter can be defined
from the long distance limit of the correlation functions. This seems to agree
with that found in the analytical work of
Ref.~[\onlinecite{almeida:78b}], and is presumably the same as the local order
parameter in a symmetry breaking field. Hence, there seems to be no
inconsistency if the limit $N \to \infty$ is taken first.

We have also studied the $m=\infty$ model in three dimensions, finding the
transition to be at zero temperature, in contrast to the situation
for~\cite{ballesterosetal:00,lee:03}
$m = 1, 2$ and $3$. We suspect that $T_{SG} = 0$ only in the 
$m=\infty$ limit,
rather than for all $m$ less than some (non-zero) critical value
$m_c$, since spin glasses with $m = \infty$ seem to have 
unique features. We have already mentioned that there is only quasi long-range
order below $T_{SG}$ in this case, in contrast to finite-$m$.
Another example is that Green et al.~\cite{green:82} find the upper
critical dimension, above which the critical exponents are mean field like, 
to be $d_u = 8$, whereas for finite $m$ one has $d_u = 6$.
Our result that $T_{SG} =
0$ for $m=\infty$ in $d=3$ is \textit{consistent} with the claim of
Viana~\cite{viana:88} that 
that the lower critical dimension (below which $T_{SG}=0$)
is also $d_l =8$, but currently we cannot say anything specific about
dimensions above 3.

We find, not surprisingly, that
$T_{SG}=0$ also in two dimensions. Our results for the correlations length
exponent at the $T=0$ transition in $d=2$ and 3 are consistent with those of
Morris et al\cite{morris:86}.

Finally, we note that Aspelmeier
and Moore~\cite{aspelmeier:04} have proposed that the
$m=\infty$ model is a better starting point for describing Ising or
Heisenberg
spin glasses in finite dimensions, than the Ising model. We have argued in
this paper that the spin glass with $m$ strictly infinite is not a sensible
model, but one rather needs to consider $m$ large but finite. Hence, the
$m=\infty$ formulation proposed by
Aspelmeier and Moore~\cite{aspelmeier:04} and
others~\cite{bray:82,morris:86,hastings:00} would need to be extended to a
$1/m$ expansion and evaluated, at the very least, to order $1/m$. More
probably an infinite resummation would be needed (M.~A.~Moore, private
communication) to obtain sensible results in the spin glass phase, but this
may be feasible. 

\begin{acknowledgments}
We acknowledge support
from the National Science Foundation under grant DMR No.~DMR 0337049. We would
like to thank Mike Moore for helpful correspondence on an earlier version of
this manuscript.

\end{acknowledgments}

\bibliography{refs}

\end{document}